# Condensed Past, Thick Present: Evolutionary Approach to the Conscious Experience

Anna Sverdlik

Sheba Medical Center, Tel-Aviv University

Email: Anna.Sverdlik@sheba.health.gov.il

## Abstract

This paper examines the conceptual convergence between Lee Smolin's *Causal Theory of Views*, Karl Friston's Free Energy Principle, and contemporary psychological accounts of the functions of consciousness. Although formulated within different domains—physics, biology, and psychology—all three frameworks, in one form or another, appeal to processes of transition from uncertainty to certainty, in which novelty arises through the resolution of surprise. According to the first two approaches, these transitions are realized within particular temporo-spatial gaps, which themselves evolve and become increasingly elaborate as organization grows. By tracing the structural and functional parallels between these frameworks, the paper proposes an account of how the evolution and the gradual elaboration of novelty, surprise, and these temporo-spatial gaps may be linked to the emergence and progressive development of consciousness, up to its highest forms addressed by psychology.

## Introduction

There has long been a need for a theory that treats the world as a unity of the inanimate and the living, the objective and the subjective. This problem is often framed as the problem of consciousness. Resolving it requires joint efforts from very different fields, from mathematicians and theoretical physicists to psychoanalysts. They share no common language, and there is concern that this project will end up like the Tower of Babel.

The main initiative in creating a theory that would encompass subjective phenomena, including consciousness, has come from physicists. Their interest is driven, first of all, by practical reasons, the best-known of which is the observer paradox: the collapse of a particle's wave function upon observation or measurement, characterized by a transition from quantum-mechanical to classical behavior. Until the observer paradox is resolved, no consensus interpretation of quantum mechanics is in sight. Hence, physicists start from the bottom, seeking the rudiments of consciousness in the inorganic world, beginning with quantum-mechanical phenomena. Beyond practical considerations for many physicists this is a problem of an even more fundamental nature: a theory that ignores such an obvious influence of the living on the physical world—that, in the end, ignores us—is incomplete and cannot be true. One need not appeal to quantum mechanics to see this: every building around us—an indisputable physical reality—once existed solely as an idea; without a subjective intention to bring it into being, it would never have appeared. No existing physical theory explains that intention or how an intention could become a building.

# Subject and Consciousness within Neuropsychology

The boundaries of the concepts "subject" and "consciousness" are extremely hard to define since they reach far beyond what lanquage can express. In addition, they are used in very different contexts and to describe very different phenomena—from pure awareness of one's own existence to reflective and abstract conscious thinking unique to human beings. For that reason even in a specific context it is impossible to provide a precise definition; the best one can do is convey the meaning of "consciousness" by describing its functions.

According to neuropsychology, one of the central functions of consciousness is to evaluate situations for which there is no precedent and no ready-made solutions, that is, to resolve situations that are marked by ambivalence and uncertainty. One might say that one of the chief functions of consciousness is the transition from a state of uncertainty to a state of certainty. We may call situations of this kind "surprisal situations," or simply surprises, which in essence they are. A surprise must be resolved definitively, because the decision must be realized as a determinate motor act: it is impossible to go both right and left at the same time.

The very phrase "transition from a state of uncertainty to a state of certainty" belongs to Lee Smolin, and the term "surprise/surprisal" to Karl Friston, whose theories this article discusses. Yet the underlying idea, as we shall see, is universal.

Put simply, we engage consciousness only when we are forced to, in situations we do not understand. In all other cases, which constitute the vast majority, we act without thinking.

In neuropsychology, the function of consciousness, namely, resolving uncertainty, is usually ascribed to the cerebral cortex, while the functions of routine, automatic thinking are assigned to subcortical structures. Decisions worked out by consciousness, i.e. by the cortex, are passed on to the subcortical regions, and after that such tasks are resolved automatically, without deliberation.

This functional definition, besides being reasonably clear, has the added advantage of allowing us (at least hypothetically) to attribute consciousness to all organisms that possess a cortex, even a rudimentary one, and not only to humans. Along with mammals, this would include birds, amphibians, fish, and possibly even cephalopods.

If we turn to psychology more broadly, **the following points are widely accepted**:
1. Consciousness is subjective; its bearer is the subject alone.
2. The subject and its consciousness are unique: each has its own view of the world, and no two views are identical.
3. The subject, and particularly its consciousness, cannot be fully understood or accessed from the outside. No external observer, whether an experimenter, a psychotherapist, or anyone else, is entirely objective. Moreover, no interaction between them is purely passive; it always affects both.
4. Consciousness endows the subject with a degree of freedom, that is, with an active capacity to resolve alternatives, or to make decisions.

# Statements

As Pauli once remarked, "a unified symbolic psychophysical language must in the end be adequate… I am quite certain that the final result will be the same, whether we begin with psyche (idea) or with physis (matter)"[1].To achieve that result, however, it may be best to begin from both ends at once and try to bring them together.

**The premises and starting points are as follows:**
1. This issue cannot be considered apart from an evolutionary perspective.
2. The evolution of the living, the subjective, and the conscious cannot be separated from the evolution of the physical world that brought them into being and ensures their continued existence and development.
3. Presumably there exists some fundamental process or a tightly connected group of processes, that pervades every level of organization and serves as the common denominator on which our understanding may be grounded. The search for this common denominator is the aim of this article.
4. The article adopts interpretations of quantum mechanics that regard the universe as nonlocal and reject the notion of an absolutely objective observer. Observation, particularly measurement, and the observed are treated as a single system, with each element influencing the other. A pure measurement does not exist. It is only a special case of interaction between a system and the world. Along with Lee Smolin, the idea of the absence of an absolutely objective observer has been discussed and developed by various physicists and philosophers of science, among them C. Rovelli[2], J. A. Wheeler[3], A. Zeilinger[4], C. Fuchs[5], and W. Zurek[6]. A similar idea is expressed by Lahav and Neemeh[7].
5. This article does not discuss the differences among various forms and levels of consciousness (though such differences undoubtedly exist) nor does it attempt to tackle the so-called "hard problem of consciousness"[8].

**The main ideas are these:**
1. The transition from uncertainty to certainty is a universal mechanism governing the existence and evolution of the inanimate and the living. It is not confined to consciousness in the narrow sense described by neuropsychology.
2. At root this process is quantum-mechanical, even when it takes the form of classical interactions and is adequately described by the formulas of classical physics. It would be more accurate to say—and a significant part of this article is devoted to the point—that the strict division into micro- and macro-world is artificial. Yet language imposes constraints on its own, and the author has no choice but to rely on the familiar dichotomous terms.
3. The passage from uncertainty to certainty, and hence the evolution of both the inanimate and the living, is bound up with the emergence of surprises: new, unprecedented events. Thus, the definition of surprise adopted in neuropsychology proves applicable to everything. Surprises act as catalysts for further, higher-ordered surprises, driving the universe toward ever-growing complexity.
4. The rise and elaboration of consciousness are inseparable from these processes. Consciousness, attributed by neuropsychology only to organisms with a cortex, is the product of a universal, sequential, continuous, and cumulative process rooted in the evolution of the microworld.
5. At the core of the transition from uncertainty to certainty in general, and of consciousness in particular, lies a temporo-spatial gap. Initially, the spatial characteristics of this gap are extremely diffuse and nearly indeterminate, but as matter becomes more complex it acquires increasingly localized and elaborate forms.
6. The physical substrates that constitute the gap may be of very different kinds. Yet, as noted in point 2, at the most fundamental level the nature of this transition is quantum-mechanical.

With this aim in mind, two ideas will be examined that may at first appear far apart. One concerns Lee Smolin's *Causal Theory of Views*[9,10,11] and, building on it, the theory of the *Physical*

*Correlates of Consciousness* he developed together with Marina Cortês[12]. The second is Karl Friston's framework, known as *Predictive Coding under the Free Energy Principle*[13,14,15]. Both hypotheses are complex and multifaceted, but in this article they will be presented only in the context relevant to our discussion.

**The article is structured as follows:**
1. The approach of Lee Smolin and Marina Cortês to the physical foundation of subjectivity and consciousness.
2. Karl Friston's *Free Energy Principle*.
3. Points of convergence and correspondence between the two theories.
4. A resulting rethinking of certain physical processes associated with subjectivity and consciousness, and of their evolutionary pathways.

## Ideas of Lee Smolin

Lee Smolin's framework, known as the *Causal Theory of Views*, stems from his work as a physicist widely recognized for contributions to quantum physics, loop quantum gravity, string theory, and cosmology. As his publications show, the problem of qualia and consciousness has preoccupied Smolin since the mid-1970s. In his opinion, any version of quantum mechanics that ignores subjective phenomena is fundamentally incomplete. In recent years Smolin, together with Marina Cortês, has developed the idea of quantum-mechanical precursors of consciousness building on his framework. This article draws on their recent work, especially Smolin's "On the Place of Qualia in a Relational Universe" (2020)[16]. Although their ideas remain well outside the mainstream, they appear exceptionally interesting and fruitful.

Within this article, building on their joint framework, Smolin offers his perspective on the emergence and development of the correlates of consciousness, beginning with quantum-mechanical phenomena. As a physicist, Smolin devotes greater attention to processes in the microworld, but he emphasizes that the general principle remains universal.

Here are **the propositions of the *Causal Theory of Views* relevant to our discussion:**
1. The universe is closed, nonlocal and asymmetric. Its fundamental characteristics are energy–momentum, causality, time, and event.
2. In this article, the primary focus is the event. Smolin defines an event as a transition from uncertainty to certainty.
3. Events form causal chains in which prior events (parents) interact with one another to generate subsequent events (children)[17]. For convenience, we label the causal level of parents as *n*, and the level of their children as *n+1*[18].
4. Not every event that has parental potential realizes it in practice: there is an element of chance in the selection of the parents. Thus, in the transition from one generation to the next there exists a tiny temporal gap to which a certain indeterminacy is inherent. This gap allows the random process to deviate from the usual course of things, which grants the universe a certain degree of freedom. Smolin refers to this temporal gap as the "thick present."
5. Children receive information about their past exclusively from their parents, and in no other way. They have no direct access to the past that preceded the emergence of their parents. One might say that all information about their past is condensed in the parental level.

6. The event (the transition from uncertainty to certainty) is the only thing that exists ontologically. The past is fixed once and for all and is represented solely by the parental level; the future is undetermined. Only the transition from uncertainty to certainty—the thick present—constitutes actual existence. Thus, the conventional distinction between past, present, and future is essentially abolished. There is only a one-way cascade of events: transitions from uncertainty to certainty. One generation of events replaces another, forming the fabric of time, but for what exists actually, time is constantly reset. Hence, the thick present, the only ontological reality, exists and is realized in unmediated form only in the gap between two adjacent generations (levels).
7. Any event, through its interaction with other events, acquires a certain representation or knowledge of the universe and, consequently, of its own place within it (bearing in mind that, in the microworld, locality is absent and "position" denotes rather a set of abstract relations). Smolin calls this knowledge the event's view. The universe is the totality of such views. What is at issue is an event's *own view from within, not* its being viewed *from the outside*. This is of fundamental importance, for it assigns subjectivity a primary ontological status.
8. An event's view is determined by two factors: its temporal history and its spatial environment. Two events with highly similar pasts have very close views; the same applies to neighboring events that share an environment. The first variant prevails in the microworld, the second in the macroworld. Nevertheless, both factors exert some influence on all scales.
9. Two identical events and, accordingly, identical views, do not exist: they are energetically forbidden. However, the overwhelming majority of quantum-mechanical events are almost or practically identical: very close copies of one another. In other words, determinism prevails in the sequence of such routine events, and they reproduce one another with almost no variation.
10. Yet at times in the course of generational succession, events or views that deviate significantly from the routine arise by chance. They are unique: they occur for the first time and have no close copies. For the sake of terminological consistency, let us call them surprisal events, or simply surprises.
11. Surprisal events, unlike routine ones, not only realize the freedom of choice afforded by chance but, by their very existence, also increase the degree of freedom of the universe as a whole. Their interactions with routine events, as well as with one another, raise the probability and accelerate the rate of generating additional surprises. This process is cumulative: as a result, the universe becomes ever more complex and diverse, gradually evolving toward the macroworld. Nothing prevents mutual influence among surprises at different levels and scales, nor their influence on other elements of the universe.
12. It must be emphasized that the extent to which a surprise influences the universe and catalyzes its further development is determined by its complexity. Naturally, macroscopic objects are more surprisal in this respect, yet even among them the differences are enormous: a human being is incomparably more of a surprise than a stone of roughly the same weight.
13. According to Smolin, surprisal events are physical correlates of conscious perceptions (PCCs). He describes this idea as a form of restricted panpsychism, which, unlike classical panpsychism, attributes certain correlates of consciousness not to everything but only to specific physical processes.

14. Neuroscience investigates neural correlates of consciousness (NCCs). This research is unquestionably necessary and important. At the same time, it should be remembered that neural correlates necessarily involve physical processes. Therefore, elucidating the nature of the physical correlates of consciousness should enrich neuroscience, as well as physics as a whole, since without understanding the phenomena of consciousness, physics cannot be complete.
15. By his own admission, Smolin's theory does not claim to resolve the problem of the dualism between matter and consciousness. He speaks specifically of correlates rather than a physical basis for consciousness, and he deliberately avoids definitions that would equate the physical with the mental.

Let us outline **the key points and their interpretation:**
1. The universe contains an element of indeterminism that may be realized in the transition from one quantum-mechanical generation of events to the next. When this happens, it gives rise to unique, or surprisal, events—those that differ substantially from routine ones. A surprisal event is a certain salience, an irregularity, a kind of turbulence in the ocean of the universe's dreary uniformity. A surprise, itself the product of chance, catalyzes the emergence of further surprises. As a result, the universe becomes increasingly complex at an accelerating pace, and part of it evolves toward the macroworld.
2. A fundamental element of the universe is the event and its view: the totality of all the relations within the universe in which the event takes part. The view of a given event belongs *to it alone*, to nothing and no one else. According to this definition, even at the most elementary physical level we are already dealing with a potential for subjectivity. Surprisal events, unlike routine ones, realize this potential: beyond possessing a unique view, they exert a disproportionately large and ever-increasing influence on the evolution of the universe. Here we find the chief characteristics that classical psychology ascribes to the subject and consciousness: (1) uniqueness and a unique view of the world; (2) salience—separation from the surrounding world—combined with active influence upon it; and (3) the capacity to alter the routine course of things and introduce an element of novelty into one's own existence and into external reality.
3. For each subsequent quantum-mechanical generation of events, the entire preceding history is homogenized and condensed within the generation of its parents. Only the transition from the preceding generation of events to the next—the transition from uncertainty to certainty within the instantaneous temporal gap—possesses ontological existence. Smolin refers to this as the thick present. The process of this transition is unidirectional and irreversible.

# Framework of Karl Friston

Karl Friston's theory, known as *Predictive Coding under the Free Energy Principle*, emerged in 2006. Since then it has developed rapidly and gained numerous adherents. Today it stands as one of the leading frameworks for understanding the phenomena of life and thought in the broadest sense. The theory has proved especially appealing to researchers in neuroscience and related disciplines and is therefore often presented as explaining the functioning of the nervous system and the brain—implying by default that its scope is limited to them. In reality, however, Friston, as a physicist, psychologist, neurophysiologist, neuroradiologist, and psychiatrist, has succeeded in integrating his broad expertise into what is essentially a unified theory of the living, attracting scholars from a wide range of fields.

The principal tool of Friston's theory is statistical physics. Like Smolin's framework, it is complex and multifaceted. I will not go into its details here, but only highlight **the points relevant to our discussion:**

1. The primary condition for the existence of life is the reduction of entropy. An increase in entropy beyond certain limits is incompatible with life.
2. Energy in a living organism exists in two forms: bound and free. Bound energy serves the organism's functions, performing useful work and reducing entropy. In contrast, free energy is not used for any purposeful process and is a source of rising entropy; it carries destructive potential. In its most fundamental form, free energy is thermodynamic energy.
3. The reduction of free energy is of utmost importance for the organism. Yet, no matter how it strives to eliminate free energy, this can never be achieved completely: according to the second law of thermodynamics, entropy inevitably increases.
4. One of the main sources of free energy is uncertainty brought into the organism by unfamiliar, novel, or unrecognized signals - in other words, surprisal signals. A surprise is a signal that deviates from the expected, routine one, the prediction. More precisely, it is not the signal itself but the difference between what was expected and what actually occurred, once that difference exceeds certain bounds. Those bounds are set by prior experience, and a surprise is, accordingly, what lies outside them. The expected, that is, what falls within the limits of prior experience, is called a concept, while what is actually experienced in real time is called a percept.
5. For most people, the terms "percept" and "concept" are automatically associated with the nervous system and the brain. It should be emphasized, however, that in this context they apply to any living organism, including even a single-celled one. For example, a change in ambient temperature detected by the cell membrane is a percept, whereas the internal model of how to respond to that change is a concept.
6. Prior experience is never represented in its entirety. What falls within a concept is an abstracted model that integrates similar elements of prior experience. The concept is based on the statistics of past events, or, more precisely, on the mechanisms of Bayesian inference[19]. In essence, it corresponds to the homogenized, condensed past described by Smolin, as applied to living systems. In living beings, this past is represented only as the bounds of expectation—and nothing more.
7. A surprise compels the organism to take action. First, it must verify the surprise, for instance by examining an unfamiliar object from different angles or by cross-checking facts. If the signal proves genuine and not a false alarm, the organism then assimilates it, binding the free, destructive energy introduced into the system and thereby reducing entropy.
8. A living system is organized hierarchically. Within it, the flow of signals proceeds in parallel both from the periphery to the center (bottom-up) and from the center to the periphery (top-down). The term "periphery" usually refers to those aspects of the organism that are in direct contact with the external world and thus serve as intermediaries between the organism and its environment: for example, the sensory organs in an animal, or the cell membrane in a unicellular organism. The term "center" usually refers to the structures and functions that perform a regulatory and integrative role: based on information received from the periphery, they issue downward commands that determine what to do in a given situation. In mammals, this role is played by the cerebral cortex; in a unicellular organism, by nuclear DNA. In any case, within a living system the center and the environment never communicate directly. Their interaction is mediated by multiple intermediate levels (or layers), forming a hierarchical structure in which each level can signal only to the one immediately above it. The notions of periphery and center apply to any of these levels: for level $n$, level $n+1$ is its local center, while for level $n+1$, level $n$ is its local periphery. Likewise, a signal originating from level $n$ is a percept for level $n+1$, while the bounds of expectation of level $n+1$ constitute its concept. Accordingly, a surprise is a signal that lies outside those bounds, that is, beyond the concept of level $n+1$.

9. There are many pathways that transmit signals about the same object or process. They run in parallel but communicate with one another only indirectly. For example, vision and hearing each provides information about the same source, and within each modality there are additional nuances—color, direction of movement, timbre, amplitude, and so on.

10. A signal carrying information from the most peripheral levels of the organism to the most central ones is evaluated and filtered at each intermediate level. The gap between a given level and the one above it is the only place where the direct, unmediated evaluation of the signal takes place. If the signal is predictable or routine for the higher level, and for it alone, then it is bound and extinguished, meaning that the problem of free energy has been successfully resolved at that stage. If, however, the signal is perceived by level $n+1$ as surprisal, it is modified and passed on to level $n+2$, which must then determine whether it also constitutes a surprise and whether to transmit it further. A surprise, to repeat, is what the given level is unprepared for, what falls outside its prior statistical experience. Put differently: a signal originating from level $n$, or percept, embodies the immediate present for level $n+1$, while the information represented at level $n+1$, or concept, embodies that level's condensed past. Such subordination resembles a military chain of command: one may submit a report only to one's immediate superior, who alone decides whether to pass it higher. In this stepwise process of transmission, the overwhelming majority of signals are extinguished before reaching the higher levels.

11. To understand how signals are transmitted from the center to the periphery, one only needs to invert the picture just described: the top and bottom exchange places. This may create some ambiguity in describing the process, but the principle remains the same. The lower levels are tasked with excitation and reporting problems, while the higher levels are responsible for suppressing excitation, or, if suppression proves impossible, reorganizing themselves to resolve the problem. The result of such reorganization is that similar problems will no longer need to be solved in the future; the solution will already be in place. Yet again, "lower" and "higher" do not necessarily refer to, say, the retina and the cerebral cortex, since the great majority of problems are resolved at lower levels.

12. In each particular chain, a signal is unidirectional—either from the periphery to the center or vice versa. Between the ascending and descending chains there are also horizontal connections that are not direct but collateral.

13. There are essentially three possible outcomes between levels $n$ and $n+1$. First, the signal falls within the limits of expectation and is successfully bound, that is, extinguished in the gap. It does not pass without effect; rather, it reinforces the existing statistics of level $n+1$ with one more instance. But its additional effect is minor, and the richer the prior statistical experience, the smaller it becomes, so only a small amount of energy is expended. The second outcome occurs when the signal goes far beyond the bounds of expectation, that is, when it is an unmistakable surprise. In this case, level $n+1$ has nothing to decide: the signal is immediately passed upward to level $n+2$. One could say that $n+1$ simply passes the problem upward. Here again, little energy is used. But there is yet another outcome: level *n+1* receives from *n* a borderline signal, one of *doubtful* character. It then finds it difficult to determine whether the signal qualifies as a surprise. In this case, level *n+1* collaterally informs its counterparts in parallel descending chains for joint assessment. They too may hesitate and send queries to other neighboring levels. This is the most energy-consuming scenario, as it engages many more levels and generates far greater perturbations in the system. It is no accident that doubt is said to gnaw at us: it consumes a great deal of both energy and matter. In other words, the most problematic and energy-intensive moment occurs at the resolution of uncertainty at the juncture between two adjacent levels.

14. If, however, the signal is finally recognized as a genuine surprise, it is redirected along descending pathways to resolve the problem. Such a switch from ascending to descending pathways occurs, in the vast majority of cases, at relatively low levels of *n*, where the signal is likewise extinguished.

15. One cannot say that decisions are made collegially, or that the effort to have a surprise recognized as such takes place on equal footing. As already noted, a biological system, like an army or a state, is generally organized so that higher levels dominate lower ones and tend to ignore signals from below. This dominance of center over periphery is especially evident in organisms with highly developed nervous systems, where descending neurons greatly outnumber ascending ones. The concept must be more stable and reliable than the percept. This is logical and of great evolutionary significance, since both perception and action must be checked against prior evolutionary and individual experience. An organism neither can nor needs to react fully to every signal if the problem can be resolved locally. Virtually all internal physiological mechanisms normally function in this way. For instance, to digest unfamiliar food, the stomach usually requires no active intervention from the brain or from most other organs. We also literally do not see or hear most of what goes on around us: if a signal contains nothing new or relevant, it is extinguished at the earliest stages of perception or cut off altogether. Accumulated experience must weigh far more than a single surprise, unless that surprise is traumatic or fairly unusual. Only then are the uppermost levels engaged, the highest of them being conscious thought, the most valuable and most costly product of evolution. Yet even it must evaluate situations with balance, relying on the statistics of prior experience. If, for example, you spot a dinosaur across the street, do not run off in panic or conclude that dinosaurs still live among us. First, take a closer look: perhaps you imagined it. If the dinosaur is still there, you might begin to doubt that dinosaurs truly went extinct, but at first it would be wiser not to share such doubts with anyone.

16. Surprises are often unpleasant and at times even fatal for the individual. Overall, however, a system organized in this way is highly effective for the survival and adaptation of the species. On the one hand, it is low-cost; on the other, the struggle with surprises brings about renewal and modernization of the organism in response to new realities. A surprise, by engaging a particular segment of the system, trains that segment to respond to it in the future: the organism now has a plan (a concept), and subsequently the same or a similar signal will be less of a surprise or simply a routine event. Its processing will require far less energy. Since the organism drows on a common energy pool, it benefits as a whole.

17. The preparation of plans for the future, or the routinization of surprises, is the goal of any biological system as well as of all social institutions. Fire drills, rules of etiquette, and the like: once learned, they allow automatic action without wasting time and energy on doubt.

18. The more unusual a surprise, the more levels it crosses and the higher the strata it engages to resolve the uncertainty. At higher levels, resources are greater, since directives sent downward must affect an increasing number of lower links and collateral pathways, mobilizing more of the system to deal with the problem. At the very top, the mechanisms that give rise to conscious awareness this problem requires resolution become evident. This is precisely what is meant by conscious thought—the work of consciousness in neuropsychology and in the commonly accepted sense of the term. The flip side of this process is its cost in time and resources: the higher the levels engaged in resolving uncertainty, the longer uncertainty persists at lower levels, and the greater the energy cost.
19. Why are we forced to think consciously? At the very least, to resolve uncertainties so serious that they cannot be settled automatically. Surprises of this kind traverse multiple ascending and descending pathways, creating energetic conflicts between subsystems so intense that they may threaten to paralyze the entire organism. When both hunger and fatigue are equally strong—when hunger prevents sleep and fatigue prevents eating—the organism faces energetic collapse from both sides, and no reasonable compromise is possible. Mutually exclusive concepts struggle with each other at the highest level, keeping all subordinate levels under tension. To avoid falling asleep while eating, one must forcibly sustain motor activity—for example, by making oneself chew—until the body's glucose level rises above the critical threshold. This, presumably, is the physiological basis of consciousness: the most energy-intensive of all processes yet not fundamentally different from others occurring within the living system.
20. It now becomes clearer how and why surprisal events generate free energy, and how that energy is bound. Friston offers a single, elegant answer to both questions. Interaction with a surprisal signal requires far more physical work than with a routine one. The free energy it introduces is filtered, transformed and relayed across levels, modifying each segment it passes through and adapting it to similar signals. In the vast majority of cases, these processes unfold without the involvement of consciousness.
21. As a result, the organism is renewed and refined, becoming a surprise to the surrounding world and introducing changes never seen before.

**To summarize:**
1. Despite the great diversity of living organisms and their physical, physiological and psychological traits, and their modes of adaptation, the fundamental locus for evaluating and transforming a signal arriving from the periphery lies at the interface between two adjacent levels of the hierarchy. The fate of the signal is determined immediately only within the gap between them; all other influences are mediated. As Jakob Hohwy, one of the best-known researchers and popularizers of Friston's theory, writes:
   > This constellation of prediction error mechanisms then forms the perceptual hierarchy, which maintains a rich representation of the world. To achieve this, each level of the perceptual hierarchy in the brain does not need to "attempt" to represent the world, nor does it need to be supervised by itself or by external perceivers. It just needs to minimize prediction error relative to the level below it... What we have seen is that all these parts are directly motivated by the imperative to minimize error; none of the parts need to appeal to different, independent principles. We can stick to just the one rule. This extreme parsimony is one of the greatest attractions about the whole framework[20].

2. Only a signal that falls outside the expectations of the subsequent level, that is, a surprisal signal, has a chance to cross the barrier and reach the next level. In this way, the organism must expend a significant amount of bound energy to bind free energy.
3. As described above, every level is influenced from both directions: bottom-up and top-down. The bottom-up influence (percept) is immediate and direct; it constitutes the present, the very moment the organism is living through. The top-down influence (concept) is shaped by the prior statistical experience of the individual and the species. The experience accumulated at each descending link exerts pressure on the one below, transmitting its influence downward through the chain. What descends from above is a homogenized, condensed past, a seemingly fixed routine of existence. This routine is specific to each level, yet it is invariably determined by that level's own history. It changes only slowly; it is inert, as it must be.
4. A routine, familiar signal is low-cost; it is easy for the system to process, but it does little to advance it. By contrast, a surprise is something genuinely novel, unseen and unforeseen. It can alter the system, turn it into something it has never been, and make it a new source of further surprises.
5. The functioning and adaptation of any organism amount to a transition from uncertainty (surprise) to certainty (the routinization of surprise). As already noted, this process takes place *immediately* in the gap between two adjacent layers, and *nowhere else*. All other processes related to the routinization of surprise are mediated.
6. $n$ can take on any value. However, the higher the $n$, the more costly the process becomes, as the state of uncertainty is prolonged and extends over larger domains.
7. Solving a problem means resolving doubt (or uncertainty, or surprise) through the involvement of consciousness. Such problems range from the simplest daily choices to the most complex mathematical proofs. Every act of conscious thought is a working-through of some version of the sleep-or-eat dilemma. What sets consciousness apart is that the surprise reaches the highest levels of the hierarchy, where the uncertainty it brings is sustained the longest and at the greatest cost.
8. However, the function of resolving uncertainty is by *no means unique to consciousness*. On the contrary, (1) it is a universal process that, in the overwhelming majority of cases, occurs without consciousness; and (2) the characteristics of consciousness described here are quantitative rather than qualitative.

This account emphasizes that the fundamental mechanisms of consciousness are universal across life. They are found above all in the gap, in the interaction of $n+1$ with the surprise that $n$ springs on it.

## Similarities Between the Frameworks of Smolin and Friston

1. In both frameworks, the central focus is the transition from uncertainty to certainty.
2. In both, uncertainty is represented by surprise—a significant deviation from the homogenized statistics of past events. Surprise is a fundamental concept: for Smolin, it is essential to the existence and evolution of the universe in all its manifestations; for Friston, to the existence and evolution of all living things. In this sense, surprise may be treated as an attractor that effectively reduces the system's overall chaoticity, while at the same time refining itself, transforming the system's other attractors, and giving rise to new ones[21].

3. In both, the transition from uncertainty emerges and is realized immediately only in the gap between two adjacent levels *n* and *n+1*. All other influences are mediated.
4. One of the main functions of consciousness is to resolve uncertainty in those rare cases when it reaches the highest levels of n, the levels associated with conscious thought. Consciousness can thus be regarded as an extreme, yet still *particular, case of the routinization of surprise*. This is, of course, a reductionist view: consciousness is not exhausted by this. Even so, it helps to reveal the underlying principle, the shared foundation across levels of organization.
5. The concepts of levels and gaps should be understood primarily in a temporal sense: as the transition from past to present.
6. The physical instantiation may take many forms, as may the degree of localization. In the microworld, locality is absent or extremely weak; in the macroworld, it is strong. In the latter case, levels and the gaps between them can be illustrated most concretely, for example, by the space between a book and the surface of a table or by a synapse between two neurons. Yet if one accepts Smolin's view of the universe as nonlocal, it becomes clear that we are never dealing with absolute locality but with relative epicenters of events.

## Across All Gaps, the Underlying Physical Mechanisms Remain the Same

One of the ideas stated at the beginning of this article is that, at the most fundamental level, the nature of any event, and thus of both surprisalization and routinization, is quantum-mechanical, regardless of the material characteristics of the gap in which it occurs. This claim does not necessarily follow from a comparison of Smolin's and Friston's hypotheses and is, in any case, open to legitimate doubt. If it remains unproven, the common ground between the two hypotheses may be regarded as semantic rather than substantive.
Indeed, the microworld and the macroworld, the living and the inanimate, the cell and the human being differ greatly in physical terms. Yet for the phenomenon in question, the differences are fewer than they might seem.

Objections arise concerning what the quantum and classical realms could possibly have in common, and, more specifically, what they might share with respect to the gap. The standard arguments are as follows: (1) microobjects behave differently from macroobjects—most notably, the defining property of the microworld, superposition, is entirely absent in the macroworld; (2) each realm is governed by its own mathematical apparatus; and (3) the classical world aligns with our logic and intuition, whereas the quantum world runs counter to both.

Recall, however, that all the physical properties we ascribe to macroobjects—shape, temperature, solidity, fluidity, and the like, together with the interactions among them, rest on quantum-mechanical features of their microscopic organization: the Pauli principle, van der Waals forces, hydrogen bonding, electron clouds and so forth[22,23,24,25]. Superposition, in one form or another, underlies all these mechanisms. The Pauli principle constrains it[26,27], yet it remains superposition, once again confined to the microworld. In the end, it is superposition that underlies the laws of classical mechanics: a book rests on a table without merging with it, billiard balls scatter after colliding, and tea cools in a glass.

Yet the very nature of these molecules — their individual views in Smolin's sense, their identities, and the forces that both bind them and drive their motion — is governed by quantum mechanics. These laws ultimately determine the processes within the glass. Still, when such fine resolution is unnecessary, the details can be averaged away. The result is the laws and formulas of thermodynamics, in which quantum effects appear to be absent on the surface yet remain implicitly embedded. The same applies to other laws of what we call classical physics. Still, even in the macroworld there are phenomena such as superfluidity, superconductivity, or laser emission, which can be explained and described mathematically only by quantum mechanics. At the very least, for this reason, dividing the world into micro and macro looks contrived. What we are dealing with is not two fundamentally different realms but our own limited needs, further constrained by our cognitive capacities.

Generally speaking, the question should be posed as follows: are there any interactions in the classical macroworld whose foundations are entirely independent of quantum-mechanical processes? Or even a single physical quantity in the equations of classical physics that is not, at least implicitly, rooted in them? This question has been raised repeatedly, yet, as far as I know, no convincing example has ever been produced.

The dichotomy between the micro- and macroworld looks even more questionable if we consider superposition in a broader sense: the coexistence of multiple potentials, of which only one is realized for us or for our instruments at any given moment. Let us show that in the macroworld, too, such potentials are ubiquitous, yet they are sustained by quantum-mechanical processes.

To begin with, recall what superposition means in the framework of quantum mechanics: a quantum system can exist in several possible states simultaneously until it interacts with the external world. Once such an interaction occurs, the superposition breaks down, and the system resolves into one of its possible states— the wave function collapses.

Some interpretations of quantum mechanics take the "external world" to mean the act of measurement and the measuring apparatus; others construe it as the world at large, treating measurement as merely one kind of interaction. Following Smolin and many others, we adopt the latter view. Yet in both cases, the post-interaction state of the system is established only through the judgment of a human observer, whose cognitive apparatus cannot register quantum effects directly. That judgment relies solely on the readout of a macroscopic instrument whose pointer, after a sequence of unobserved quantum oscillations, comes to rest at a preassigned mark expressed as a real—that is, classical—number.

Stripped of what pertains only to quantum mechanics, the definition of superposition can be stated as follows: a system can exist in several possible states at once, but only one of them is realized at any given time. One can argue that, in the macroworld, this state of affairs is not the exception but rather the rule.

Consider a piece of ice placed in a subzero environment. According to classical physics, it would be treated as an ideal crystal. In reality, however, ice is in a constant state of dynamic fluctuation between a perfect crystal, an amorphous structure, and a liquid tendency. These fluctuations are governed by quantum mechanisms, although transitions into amorphous or liquid-like phases occur only locally within the piece of ice. In general, any solid constitutes a macroscopic stabilization, yet at the microscopic level it invariably contains local fluctuations, defects, and rearrangements that disrupt its ideal structure. Such effects have been observed in liquids, dielectrics, supercooled ferromagnets, micron-scale mirrors, and superconducting circuits—the list could easily be extended. According to Penrose's Objective Reduction (OR) hypothesis[28], local microsites of destabilization spontaneously arise within the structure of macroscopic systems, but they cannot grow or coalesce because gravity destroys them instantly. Moreover, according to the Diósi–Penrose hypothesis[29], a classical object exists not in spite of quantum superposition but because of it. Its structure is not a "cancellation" of superposition but a rapidly collapsing dynamic in which superpositions locally flare up, vanish, and overlap, producing an illusion of stability. The macroworld presupposes not the absence of superposition but its perpetually suppressed potential.

Now an example from biology. Every living organism is perpetually caught in a struggle between life and death, suspended in a state of potentiality. In every atom, every molecule, cell, and system, on micro-, meso-, and macroscopic levels alike, there is an unceasing cycle of breaking and mending, of births and burials. The question of whether an organism is alive or not is, once again, left to a classical instrument. Yet even its pointer, before finally coming to rest, flickers chaotically across the scale, on timescales that are long and clearly perceptible, in contrast to the fleeting durations of the microworld.

Finally, a word on superposition and the nature of thought. Nearly all of our mental activity is unconscious, with several alternative scenarios unfolding at once. The more developed a neural network is, the more capable it becomes of sustaining this process: its very purpose is to generate mutually exclusive potentials and to hold them in superposition until one of them collapses onto the narrow stage of our consciousness. Our cognitive apparatus can perceive only the stage, while the real event unfolds behind the scenes.

## Features of Gaps in Living Systems

The laws of molecular biology rest on chemistry and thermodynamics, which in turn arise from the quantum behaviour of particles. As in classical physics, in biology each level of complexity is subsumed by the next. Accordingly, everything said in the previous section applies to biology as well: the gap in living systems is a particular case of the gap in general. Regarding consciousness, I do not think its nature should be sought in any special biological mechanisms. These mechanisms, as already noted, appear to be universal. The search for macroscopic decoherence in axonal proteins, proposed by Hameroff and Penrose as an explanation of consciousness[30], seems conceptually unfounded, if only because rudiments of consciousness must have emerged long before the appearance of axons.

However, let us return to a question that, to my knowledge, remains unanswered: are there processes in the classical world that are truly independent of quantum mechanics in essence rather than merely in form? Is there any experimental basis for maintaining a strict micro–macro dichotomy? Applied to our topic, does the physical scale of the gap influence the fundamental mechanisms of surprisal? After all, electron clouds are one thing, while interactions between DNA and RNA, between blood and tissue, or events within a synapse between two neurons, phenomena usually assigned to the meso- or macroscopic scales, are quite another.

Consider two possibilities: it makes no difference, or it does. If it makes no difference, the common denominator in Smolin's and Friston's hypotheses is clear. If it does, however, then the role of quantum-mechanical processes in the mechanisms of surprisal and consciousness in living beings, while not absolute, is nonetheless essential—and, most importantly, *far greater than in the nonliving*.

We may now turn to the universal mechanisms that shape level *n+1* in the course of evolution—that is, through the increasing complexity of surprisal across levels, from molecular to cellular and beyond, as *n* increases within a single organism and across timescales up to the phylogenetic.

To specify the question: the point is not the specific characteristics of level *n+1* itself, but its modification—the shift from what it was to what it becomes as a result of interaction with the signal originating from level *n*. This shift, in turn, is grounded in the difference between *n+1* and the incoming signal from *n*.

Even in the simplest organisms, signals and receptors show great diversity. Yet evolution has ensured that each signal from level *n* and its receptor at level *n+1* correspond both structurally and functionally. A specific protein and its cognate receptor are like a ball and a billiard pocket. This inherent complementarity enables contact, but the outcome—whether the ball falls into the pocket or not—depends on highly specific, localized interactions between narrow loci of the protein and its receptor. Nevertheless, all these interactions follow the same fundamental physico-chemical laws, in which hydrogen bonds and electronic forces play the decisive role. For this reason, it is more appropriate to speak not of the multitude of signals and receptors but of the universal processes that determine the outcome of such interactions at every level of *n*. Thus we may ask: what are the universal physical mechanisms that assess and, when necessary, reduce the mismatch between signal and receptor? Could the processes described by Smolin play a decisive role here, or at least carry significant weight? (It should be emphasized once again that this does not imply that the physico-mathematical formulation of such processes must be quantum-mechanical. Quantum-mechanical surprisal may well be implicit in the equations of classical physics.)

In this context, it is worth recalling the quantum Zeno effect (QZE)—a phenomenon in quantum mechanics whereby frequent measurements suppress a system's transition away from its initial state, effectively slowing its evolution[31,32]. In other words, repeated measurement enhances the influence of quantum-mechanical interactions while diminishing that of classical ones.

A system's state can be measured externally (as in a laboratory experiment) or internally, through self-monitoring. [33,34]. The more complex the system and the greater its degrees of freedom, the more self-monitoring it requires. Biological systems are evidently more complex and engage in self-monitoring and self-regulation to a much greater extent than nonliving ones. Let us now consider how self-monitoring may be related to the Zeno effect.

Recognizing a signal is a kind of appraisal, an act of measuring it by level *n+1*. If the signal is routine and fits expectations, a single measurement, or at most a few, should suffice. If level *n+1* registers uncertainty, it should recheck the signal, for instance by examining it from another aspect. The prefix "re-" already implies this: to decide "yes" or "no," the signal must be measured more intensively through varied checks, a greater number of repetitions, and evidently *at a higher frequency*.

Such an assumption, though indirect, has numerous and convincing experimental confirmations. The main evidence is that when level *n+1* encounters surprise, the frequency it transmits increases sharply. This phenomenon has been observed in neuronal synapses[35], in smooth muscle cells[36], in immunocytes[37,38], and even in plant cells[39]. The most plausible explanation is increased stimulation from the underlying level.

Another important point must be kept in mind: uncertainty within one gap requires the activation of neighboring gaps for its resolution. The greater the uncertainty, the more gaps are drawn into the process, the more signals pass between them, and the more intense the rechecking within each becomes. In this way, the quantum Zeno effect spreads and amplifies as uncertainty increases, that is, as the task grows more complex. Its influence reaches a maximum at the highest levels of complexity—precisely in problems that demand the active involvement of consciousness.

The quantum Zeno effect is also facilitated by the system's substantial isolation from the environment and by a high level of entropy within the system relative to its surroundings[40,41]. Both of these conditions are met in the Fristonian gap. As noted earlier, the contact between signal and receptor is relatively isolated, both structurally and functionally. As for entropy, this point has also been emphasized more than once: the amount of work required to resolve uncertainty in the gap is at its maximum, and so too are the local temperature and local entropy.

Thus, it may be assumed that in the process of resolving uncertainty — that is, in consciousness in the broad sense — direct quantum-mechanical mechanisms play a greater role than is commonly supposed. For now, this remains a hypothesis in need of experimental verification.

## The Living and the Nonliving: Differences in the Light of the Two Frameworks

Smolin's framework emphasizes the fundamental nature of time, underscoring its inseparable connection to another of his foundational notions—the event. Let us restate his key points on this matter:
1. The transition from uncertainty to certainty corresponds to the passage from one generation of events (the "parents," $n$) to the next (the "children," $n+1$). This cascade of events is irreversible. It weaves the fabric of time; it is time itself. Space takes shapes and develops as surprises evolve, with more and more of them becoming complex and extensive enough to reach the scale of the meso- and macroworld.
2. All information about past events is statistically averaged and condensed within the parental level. The progenitors are nothing but statistics; for the children, in the ontological sense, they no longer exist.
3. If we allow for the existence of an external observer, time is unidirectional: the arrow of time moves only one way. Yet because no purely external observation is possible, and because what is fundamental is Smolin's subjective "view," time is effectively reset with each successive event. Ontological existence belongs only to the present—the thick present, as Smolin defines it. In this sense, both past and future fall away.

Building on these principles, and once again setting them against Friston's hypothesis, let us consider how the **distinction between the organic and the inorganic** may be defined:
1. In both frameworks, the passage from uncertainty to certainty arises and is realized immediately within the gap between neighboring levels or generations, $n$ and $n+1$. This local process provides the fundamental ground of subjectivity and consciousness; all other processes are mediated and secondary. The resolution of uncertainty, or of surprise, constitutes the thick present in both Smolin's and Friston's frameworks.
2. According to Friston, level $n$ sends to level $n+1$ a percept—a signal carrying information about what is happening in the present moment. This signal is the unaveraged, momentary, living present of level $n$. *This* present belongs to *n and to n alone*; it is its subjective, Smolinian view. An external observer might call it the present continuous. The signal interacts with its corresponding concept—that is, with the representation of the past provided by level $n+1$; an external observer might call this the past perfect. This past is, in a sense, dead like Smolin's past. Yet unlike Smolin's, which is dead once and for all, Friston's past does not vanish completely; it revives whenever the present collides with it, taking it by surprise.
3. Thus, for living systems, the past appears to be dual. One past is unidirectional and irreversible, as in Smolin's framework. The other exerts a retroactive force, shaping what happens here and now. This second past cuts across the present, pre-empting many of the problems that unwelcome surprises would otherwise bring.

4. This dialectic allows the organism to conserve energy in real time while continually refining the mechanisms of conservation throughout development, thereby increasing efficiency. The more effective these processes are, the greater the prospects for survival and flourishing—at the levels of proteins, mitochondria, cells, and organisms alike.

## Conclusions

This paper proposes a unified generative model that threads through both the inanimate and the living, from quantum events to conscious deliberation. It is the passage from uncertainty to certainty, through which departures from routine—the surprises—emerge and drive the progressive differentiation and increasing complexity of the world. These transitions from uncertainty to certainty are realized immediately within gaps, which themselves evolve and become more elaborate as organization grows.

In Smolin's Causal Theory of Views, the universe unfolds through successive gaps in which each event, within its thick present, converts the indeterminate into the determinate—an irreversible act of resolving surprise into a novel structure, a novel reality that becomes routinized within it and serves as the epicenter of new surprises. This continual renewal drives the evolution and increasing complexity of all that exists. Smolin conceives this transition as a universal principle, though in his discussions he focuses chiefly on its physical manifestations. Friston's free-energy framework is principally akin: it renders the same passage in biological terms. Within the organism's hierarchy of gaps, each surprise injects free energy into the system until it is gradually absorbed and routinized, leading to the organism's refinement and an increased capacity for sequential surprises. A surprise in Karl Friston's sense is, in essence, Lee Smolin's surprisal event translated into the language of life.

Thus, both theories rest on the same triad: the transition from uncertainty to certainty, the generative force of surprise, and the temporo-spatial gap as the site of that transition. Yet in living systems, this gap acquires a twofold temporal structure: there are not one but two opposite arrows of time. In addition to an immediate, living, and irreversible thick present, the living possesses a reflective, condensed past that directly interacts with the thick present in every act of becoming.

Together, the two frameworks imply that subjectivity, and in its culminating form consciousness, arises from a single universal mechanism. The line that seems to divide the living from the nonliving marks not a rupture of principle but a deepening of structure, coupling, and temporo-spatial reach—a widening of the world's own capacity for self-reference. Neural correlates of consciousness (NCCs) thus appear as one family within a broader class of physical correlates of consciousness (PCCs), extending the same principle across the scales of matter and life.

Of course, this remains a hypothesis—not to collapse psychology into physics but to outline a shared architecture: events that resolve uncertainty in local gaps and, in doing so, generate novelty. If borne out, it reframes consciousness as the most intricate fold of a single fabric of existence.

[14] Buckley, C. L., Kim, C. S., McGregor, S., & Seth, A. K. (2017). The free energy principle for action and perception: A mathematical review. Journal of Mathematical Psychology, 81, 55–79. https://doi.org/10.1016/j.jmp.2017.09.004

[15] Ramstead, M. J. D., Badcock, P. B., & Friston, K. J. (2018). Answering Schrödinger's question: A free-energy formulation. Physics of Life Reviews, 24, 1–16. https://doi.org/10.1016/j.plrev.2017.09.001

[16] Smolin, L. (2022). On the Place of Qualia in a Relational Universe. In S. Gao (Ed.), Consciousness and Quantum Mechanics. Oxford University Press.

[17] In Smolin's framework, the notion of an event is inherently dual. It denotes both the process of transition from uncertainty to certainty and the outcome of that process—the realized state that serves as a source of causation for subsequent events. This duality likely stems from Smolin's central assumption that reality consists of ongoing acts of becoming, in which each event both closes one uncertainty and opens the next.

[18] Smolin typically refers to successive sets of events as "generations" or, occasionally, "causal layers." In contrast, Friston's terminology is inherently hierarchical and employs the notion of "levels" to describe nested generative models in biological systems. Since the present article aims to integrate both frameworks, the term "level" is adopted throughout for consistency, while retaining Smolin's causal logic.

[19] Mechanisms of Bayesian inference operate by deriving conclusions from the statistical structure of prior experience. They integrate new sensory data with probabilistic expectations formed from past events, continuously updating internal models so as to minimize surprise and maintain coherence between prediction and perception.

[20] Hohwy, J. (2013). The Predictive Mind. Oxford University Press, Oxford, p. 78

[21] Lahav, N., Sendiña-Nadal, I., Hens, C., Ksherim, B., Barzel, B., Cohen, R., & Boccaletti, S. (2022). Topological synchronization of chaotic systems. Scientific Reports, 12, 2508. https://doi.org/10.1038/s41598-022-06262-z

[22] Israelachvili, J. N. (2011). Intermolecular and Surface Forces, 3rd ed.; Academic Press.

[23] Wagner, J. P., & Schreiner, P. R. (2015). London dispersion in molecular chemistry—reconsidering steric effects. Angewandte Chemie International Edition, 54(42), 12200–12207. https://doi.org/10.1002/anie.201503476

[24] Sakurai, J. J., & Napolitano, J. (2021). Modern Quantum Mechanics (3rd ed., Chapter 7: Identical Particles). Cambridge University Press. https://doi.org/10.1017/9781108587280

[25] Grabowsky, S., Genoni, A., & Bürgi, H.-B. (2022). Quantum crystallography—electron density and chemical bonding. WIREs Computational Molecular Science, e1601. https://doi.org/10.1002/wcms.1601